\documentclass[12pt]{article}
\usepackage{graphicx}
\usepackage{hyperref}
\usepackage[numbers,sort&compress]{natbib}

\usepackage{amsmath}
\usepackage{amsfonts}
\usepackage{amssymb}%

\begin{document}

\begin{center}

\hfill LMU-ASC 40/12\\
\vskip 1 cm

{\Large\bf  Semiclassical Calculation of Multiparticle Scattering Cross Sections in Classicalizing Theories}
 \vspace{0.2cm}

\end{center}

\begin{center}

{\bf Lasma Alberte$^{a,}$\footnote{Email: \texttt{lasma.alberte@physik.lmu.de}}, Fedor Bezrukov $^{b,c,}$\footnote{Email: \texttt{Fedor.Bezrukov@uconn.edu}}} 

\vskip 0.3cm

\centerline{\em $^a$Arnold Sommerfeld Center for Theoretical Physics,
Fakult\"at f\"ur Physik} 
\centerline{\em Ludwig-Maximilians-Universit\"at M\"unchen,
Theresienstr.~37, 80333 M\"unchen, Germany}

{\em $^b$Physics Department, University of Connecticut, Storrs, CT 06269-3046, USA}

{\em $^c$RIKEN-BNL Research Center, Brookhaven National Laboratory, Upton, NY 11973, USA}

\end{center}


\centerline{\bf Abstract}

It has been suggested that certain derivatively coupled non-renormalizable scalar field theories might restore the perturbative unitarity of high energy hard scatterings by classicalization, i.e.\ formation of multiparticle states of soft quanta  \cite{Dvali:2010jz}. Here we apply the semiclassical method of calculating the multiparticle production rates to the scalar Dirac-Born-Infeld (DBI) theory which is suggested to classicalize. We find that the semiclassical method is applicable for the energies in the final state above the cutoff scale of the theory $L_*^{-1}$. We encounter that the cross section of the process $2\to N$ ceases to be exponentially suppressed for the particle number in the final state $N$ smaller than a critical particle number $N_\textrm{crit}\sim (E L_*)^{4/3}$. It coincides with the typical particle number produced in two-particle collisions at high energies predicted by classicalization arguments.

\section{Introduction}
Traditional approach to the field theories proposes, that the fundamental (and allowing for predictive calculations) field theories at high energies are the renormalizable ones.\footnote{Another option may be asymptotic safety, corresponding to theories with nontrivial RG fixed point in the UV, proposed in \cite{weinberg}. However there are no reliable calculations for such theories in most cases at present.}  A non-renormalizable effective theory at lower energy may have two different kinds of behavior at high energy. It can either complete itself at UV by additional weakly coupled perturbative degrees of freedom (Wilsonian completion) and become a renormalizable theory, or it can match to a strongly coupled phase of an asymptotically free theory. A well known example for the Wilsonian UV completion is the four-fermion theory of weak interactions at low energy becoming a gauge theory with the Higgs mechanism above the Fermi scale. Alternatively, the effective theory describing baryons and mesons at low energy is completed at high energies by the asymptotically free QCD with gluons and quarks.  Recently an alternative mechanism, termed \emph{classicalization}, was suggested in \cite{Dvali:2010jz} for theories with non-renormalizable derivative self-couplings.
This mechanism may work in such fundamental theory as gravity \cite{Dvali:2010bf,Dvali:2010ue}.

The simplest example of a classicalizing theory is a scalar theory with the leading nonlinear derivative interaction of the form
\begin{equation}
  \label{eq:dphi4}
  L_*^4(\partial_\mu\phi\partial^\mu\phi)^2.
\end{equation}
A particularly convenient example of a scalar field theory with such a leading interaction term is given by the Dirac-Born-Infeld (DBI) type action
\begin{equation}\label{dbi0}
S=\epsilon_2\int d^4x\,\frac{1}{2L_*^4}\sqrt{1+2 \epsilon _2L_*^4\left(\partial_\mu\phi\right)^2}
\end{equation} 
with $\epsilon_2=\pm 1$. According to the standard picture the perturbative unitarity in such theories is violated at energies above the cutoff $L_*^{-1}$ due to the derivative self-interactions of the scalar field. Instead it was suggested in \cite{Dvali:2010jz,Dvali:2010ns} that in such theories a trans-cutoff scattering processes of two particles are dominated by low momentum transfer $\sim r_*^{-1}$, where the length scale $r_*(E)$ depends on the energy and $r_*(E)\gg L_*$. As a result the leading contribution to the scattering process of two hard particles with high center of mass energy $E\gg L_*^{-1}$ comes from the production of a multiparticle quantum state of $N\sim Er_*$ soft particles. This state is called \emph{classicalon} and in the semiclassical limit 
$$
L_*\to0,\qquad N\to \infty,\qquad r_*=\textrm{fixed}
$$
it should correspond to a classical configuration of size $r_*$, which is a solution of the theory \cite{Dvali:2012zc}. The length scale $r_*(E)$ is called the \emph{classicalization radius}. In this way the theory self-unitarizes by prohibiting to probe small distances $r\ll L_*$ in high energy scattering processes.

The focus of the present work is the semiclassical calculability of multiparticle production in such theories. For the convenience of calculations we will focus on the scalar DBI action \eqref{dbi0}. In conventional weakly coupled scalar field theories with a dimensionless coupling constant $g$ it is known that the perturbative methods fail to describe the scattering amplitudes for processes with large particle number $N$ in the final state.   This happens when the multiplicity of the final state $N$ becomes of the order of the inverse coupling constant $1/g^{2}$. Therefore, in the limit when $g\to 0$ and $N\sim 1/g ^{2}$, non-perturbative methods are used to calculate the cross sections of multiparticle productions from a few (hard) initial particles. However, in the scalar DBI action \eqref{dbi0}, which is of the main interest in the present paper, the coupling constant $L_*$ has the dimension of length. The above estimate of the critical multiplicity of the final state does not immediately generalize to theories with dimensionful couplings. One of the side results of this paper is that we modify the semcilassical technique for the calculation of the multiparticle cross sections developed in \cite{Khlebnikov:1990ue,Rubakov:1991fb,Son:1995wz,Bezrukov:1995ta,Bezrukov:1999kb} so that it can be applied also to theories with dimensionful coupling constants. We will then use this semiclassical technique to calculate the multiparticle production rates for the theory \eqref{dbi0}, which might exhibit the classicalization phenomena. This method is very similar to the method used to calculate high energy instanton-like transitions in the electroweak theory (for details see \cite{Ringwald:1989ee,Espinosa:1989qn,Khlebnikov:1990ue,Rubakov:1991fb}). Using the coherent state formalism \cite{fadeev} allows one to reduce the problem of calculating the cross section to solving a classical boundary value problem for the scalar field. A distinctive feature of the multiparticle processes from instanton transitions is that the field configuration saturating the scattering cross section is singular at the origin \cite{Son:1995wz}. This approach of singular solutions has been previously applied to the $\lambda\phi^4$ theory in \cite{Son:1995wz,Bezrukov:1995ta,Bezrukov:1999kb}. It has successfully reproduced all the results known from perturbative tree-level calculations, as well as the exponentiated part of the leading loop contributions \cite{Son:1995wz}. For a review of multiparticle processes and semiclassical analysis in generic scalar field theories see \cite{Libanov:1997nt}.

The purpose of this paper is to apply this semiclassical technique to calculate the transition rate of the process $\textrm{few}\to N$ in the scalar DBI theory, which was suggested to classicalize in \cite{Dvali:2010jz}. The paper is organized as follows. In section \ref{sec:2} we review the semiclassical method used for the calculation of the multiparticle cross sections and briefly present the previous results for the $\lambda\phi^4$ theory in section \ref{sec:3}. We apply the technique to the DBI theory in section \ref{sec:4}. We first discuss the semiclassical limit for this theory and find that to any given energy $E$ one can associate a length scale $r_*(E)$, such that it remains constant in the semiclassical limit. We show that this legth scale is $r_* = L_*(L_* E)^{1/3}$ and coincides with the classicalization radius of \cite{Dvali:2010jz,Dvali:2010ns}. We then report the results for the scattering cross section. For a fixed above-cutoff total energy $E>L_*^{-1}$ in the final state we find that the scattering processes with large number of particles in the final state $N>N_\textrm{crit}\sim (E L_*)^{4/3}$ is exponentially suppressed. For particle numbers $N<N_\textrm{crit}$ the exponent of the scattering cross section becomes positive. We thus see an emergent critical length scale $r_\textrm{crit}\equiv N_\textrm{crit}/E$ which also coincides with the classicalization radius $r_*$. We conclude in section \ref{sec:5}.

\section{Semiclassical formalism}\label{sec:2}
Here we briefly reproduce the derivation of the semiclassical approach to calculating the multiparticle production rates of \cite{Son:1995wz,Bezrukov:1995ta}. For further details on the  formalism of section \ref{sec:2.1} see \cite{Son:1995wz}, and for the section \ref{sec:2.2} see \cite{Bezrukov:1995ta}.

\subsection{Generic boundary value problem}\label{sec:2.1}

The total scattering cross section from an initial few-particle state to all possible final states with given total energy $E$ and particle number $N$ can be calculated as
\begin{equation}\label{cs0}
\sigma(E,N)=\sum_f\left|\left\langle f|P_EP_N\hat S\hat A|0\right\rangle \right|^2,
\end{equation}
where the operator $\hat A$ creates an initial state from the vacuum (see the discussion on next page), $\hat S$ is the S-matrix, and $P_E$ and $P_N$ are the projection operators to states with energy $E$ and number of particles $N$ respectively. The sum runs over all final states $|f\rangle$. By using the coherent state formalism \cite{fadeev} the equation \eqref{cs0} can be written as \cite{Khlebnikov:1990ue}
\begin{multline}\label{cs1}
  \sigma(E,N)=\int db_\textbf{k}^* db_\textbf{k} d\xi d\eta\mathcal D\phi\mathcal D\phi '\exp \Bigg(-\int d\textbf{k} b_{\textbf{k}}^* b_\textbf{k} e^{i\omega_\textbf k\xi+i\eta}+iE\xi+iN\eta\\
+B_i(0,\phi_i)+B_f(b_\textbf{k}^*,\phi_f)+B_i^*(0,\phi_i ')+B_f^*(b_\textbf{k},\phi_f ')+iS\left[\phi\right]-iS[\phi ']\\
+J\phi(0)+J\phi '(0)\Bigg)
\end{multline}
where $J$ is some arbitrary number defining the initial few-particle state as $|i\rangle\equiv\hat A|0\rangle=e^{J\phi(0)}|0\rangle$ and the boundary terms are
\begin{align*}
  B_i(0,\phi)= & -\frac{1}{2}\int d\textbf k\,\omega_\textbf k\phi_i(\textbf k)\phi_i(-\textbf{k}),\\
  B_f(b_\textbf{k}^*,\phi_f)= & -\frac{1}{2}\int d\textbf{k}\,b_\textbf{k}^* b_{-\textbf{k}}^* e^{2i\omega_\textbf k T_f}+\int d\textbf{k}\sqrt{2\omega_\textbf{k}}b_\textbf k^*\phi_f(\textbf k)e^{i\omega_\textbf k T_f}\\
    & -\frac{1}{2}\int d\textbf k\,\omega_\textbf k\phi_f(\textbf k)\phi_f(-\textbf k).
\end{align*}
Here $\omega_\textbf k=\sqrt{m^2+\textbf k^2}$, $T_f$ denotes some final moment of time, and $\phi_i(\textbf k)$ and $\phi_f(\textbf k)$ are the spatial Fourier transformations of the field in the initial and final asymptotic regions. The complex variables $b_\textbf k$ characterize a set of coherent states $\left|\{b\}\right\rangle$ which are eigenstates of the annihilation operators $\hat b_\textbf k$, i.e.\ $\hat b_\textbf k\left|\{b\}\right\rangle=b_\textbf k\left|\{b\}\right\rangle$ for all $\textbf k$. 

According to \cite{Rubakov:1991fb} the integral in \eqref{cs1} is of the saddle point type for any scalar field theory with some dimensionless coupling constant $g$ provided that the constant $J\sim 1/g$ and that under the change of variables $\phi=\Phi/g$ the action has the following property
\begin{equation}\label{con}
S(\phi,\,g)=S(\Phi/g,\,g)=\frac{1}{g^2}s(\Phi).
\end{equation}
In this case after the change of variables $\phi=\Phi/g$ and $(b,\,b^*)=1/g(\beta,\,\beta^*)$ the transition rate \eqref{cs1} takes the form
\begin{equation}\label{cs1b}
\sigma(E,N)\sim\int db_\textbf{k}^* db_\textbf{k} d\xi d\eta\mathcal D\phi\mathcal D\phi '\exp W,
\end{equation}
with $W=(1/g^2)F$ where $F$ depends on $\Phi,\,\Phi',\,\beta,\,\beta^*,\,gJ,\,g^2E,\,g^2N$, but does not explicitly depend on $g$. For the sake of clarity it is useful to define a new set of variables $j\equiv gJ,\,\varepsilon\equiv g^2E,\,n\equiv g^2N$ such that in the semiclassical limit $g\to 0$ they stay fixed. We will refer to these quantities as to \emph{semiclassical variables}. In the limit $g\to 0,\,j,\,\varepsilon,\,n=\textrm{fixed}$ the integral \eqref{cs1} can be taken in the saddle point approximation. We note here that the semiclassical parameter $g$ emerges naturally in the conventional scalar field theories with a \emph{dimensionless} coupling constant $g$. We will see in section \ref{sec:4} that this is not the case in theories with \emph{dimensionful} couplings. In such theories the semiclassical parameter has to be introduced by hand by demanding that the requirement \eqref{con} is satisfied.

Here we have to remark that the few-particle initial state is chosen to be of the form $\left|i\right\rangle=e^{J\phi(0)}\left|0\right\rangle$ in order to formally avoid the fact that an initial hard particle state is not semiclassical.\footnote{In principle, a different initial state can be chosen. However perturbative calculations for the $\lambda\phi^4$ theory suggest that different choices of the initial state do not change the exponent of the scattering cross section \cite{Libanov:1995gh}. The same is true also for generic scalar field theories with canonical kinetic terms \cite{Libanov:1996vq}. } In \cite{Rubakov:1991fb} it was suggested that the few-particle initial state can be recovered by first evaluating the integral in saddle point approximation in the limit $j\equiv gJ=\textrm{fixed}$ and then taking the limit $j\to 0$. The assertion is that in this limit one recovers an initial state with small number of particles. It is however not obvious that this limit indeed reproduces the amplitude for the process with few-particle initial state. For the $\lambda\phi^4$ theory a direct semiclassical calculation of the tree-level amplitudes and the exponentiated leading loop corrections was done in \cite{Son:1995wz}. Comparison with the perturbative calculations confirmed the hypothesis about the correct form of the initial state  (for perturbative calculations see e.g.\ \cite{Libanov:1995gh,Libanov:1994ug}). We will assume, that this is also true for the theory at hand, keeping in mind that this check should, in principle, be repeated.

Thus the dominant contribution to the scattering cross section \eqref{cs1} is given by the saddle point field configuration. The classical field equations and boundary conditions for the field $\phi$ are obtained by varying the exponent of \eqref{cs1} with respect to $\phi, \,\phi_i(\textbf k),\,\phi_f(\textbf k)$ and $b_\textbf k$. The explicit form of the boundary conditions can be found in \cite{Son:1995wz}. Here we write the boundary value problem for the scalar field $\phi$ in a simplified form
\begin{align}\label{eom1}
\frac{\delta S}{\delta \phi}&=iJ\delta ^{(4)}(x),\\\label{eom2}
\phi_i(\textbf k)&=\frac{a_{-\textbf k}^*}{\sqrt{2\omega_\textbf k}} e^{i\omega_\textbf kt}, & t\to -\infty,\\\label{eom3}
\phi_f(\textbf k)&=\frac{1}{\sqrt{2\omega_\textbf k}}\left(b_\textbf k e^{\omega_\textbf kT -\theta-i\omega_\textbf k t}+b_{-\textbf k}^* e^{i\omega_\textbf k t}\right), & t\to +\infty,
\end{align}
where we have assumed that the integration variables $\xi$ and $\eta$ in \eqref{cs1} are purely imaginary and substituted $T\equiv i\xi$ and $\theta\equiv -i\eta$ \cite{Son:1995wz}. The complex variables $a_\textbf k$ and $b_\textbf k$ characterize the spatial Fourier components of the initial and final  field asymptotics respectively. This result is independent on the exact form of the nonlinear scalar field interaction terms in the Lagrangian as long as the action satisfies the condition \eqref{con} and one can assume that nonlinearities can be neglected for asymptotic solutions in $3+1$ dimensions. 

There are two more saddle point equations obtained by the variation of the exponent in \eqref{cs1} with respect to parameters $T$ and $\theta$: 
\begin{align}\label{E}
E=\int d\textbf k\omega_\textbf k b_\textbf k^* b_\textbf k e^{\omega_\textbf kT-\theta},\\\label{n}
N=\int d\textbf k b_\textbf k^* b_\textbf k e^{\omega_\textbf kT-\theta}
\end{align}
This gives the physical interpretation of $E$ and $N$ as the energy and the number of particles in the final asymptotics. Due to the presence of a $\delta$-functional source located at the coordinate origin $x^\mu=0$ the energy of the system has a discontinuity at the point $t=0$. It can be seen easily from the boundary conditions \eqref{eom2} and \eqref{eom3} since at the times $t<0$ the field $\phi$ has only positive frequency modes and the energy vanishes, while at late times $t>0$ the energy is determined by \eqref{E}. Another expression for the energy in the final state can be obtained from the Lagrangian
\begin{equation}\label{E2}
E=\int_{t=0_+}^{t=0_-}dt\frac{d}{dt}\int d\textbf x\left(\frac{\partial \mathcal L}{\partial\dot\phi}\dot\phi-\mathcal L\right)=-iJ\int_{t=0_+}^{t=0_-}dt\dot\phi(t,0)\delta(t)=-iJ\dot\phi(0).
\end{equation}
Let us discuss the limitations of the allowed field configurations $\phi$ after taking the limit $J\to 0$. One sees that for the energy jump to stay finite in this limit, the derivative $\dot\phi(0)$ has to go to infinity. Hence the field has a singularity at the point $t=\textbf x=0$. Therefore, in order to evaluate the scattering cross section for the process $\textrm{few}\to N$ one has to find the solution for the boundary value problem \eqref{eom1}, \eqref{eom2}, and \eqref{eom3} which is singular at $x^\mu=0$ but regular elsewhere in Minkowski space-time. A more detailed discussion about the limit $J\to 0$ and the correct choice of the singular solution can be found in \cite{Son:1995wz}. Henceforth we will not mention the source $J$ anymore. We will nevertheless keep in mind that the condition that we are only looking for singular field configurations arises from the limit to the few-particle initial state which is equivalent to the limit of a vanishing source. 

As a result, the scattering cross section is saturated by the saddle point of the integral \eqref{cs1} and has the following form
\begin{equation}\label{cs2}
\sigma(E,N)\sim e^{W(E,N)},
\end{equation}
where 
\begin{equation}\label{W1}
W(E,N)=\frac{1}{g^2}F(n,\,\varepsilon)=ET-N\theta-2\textrm{Im}S[\phi].
\end{equation}
The saddle point relations between $\theta$ and $T$, and $E$ and $N$ can be obtained by variation of the exponent \eqref{W1} with respect to $T$ and $\theta$:
\begin{equation}\label{T}
2\frac{\partial \textrm{Im}S}{\partial T}=E,\quad -2\frac{\partial\textrm{Im}S}{\partial\theta}=N.
\end{equation}
Hence the problem of calculating the scattering cross section for multiparticle production \eqref{cs0} is reduced to solving the classical boundary value problem for the field $\phi$ stated in equations \eqref{eom1}--\eqref{eom3}. Due to the requirement of a few-particle initial state only the solution, singular at the origin $x^\mu=0$, needs to be considered. After substituting this solution in equation \eqref{W1} and using the equations \eqref{T} in order to eliminate the unphysical parameters $T$ and $\theta$ one arrives to an expression for the scattering cross section $\sigma(E,N)$.

\subsection{Euclidean version of the boundary value problem for tree level contributions}
\label{sec:2.2}

In general, solving \eqref{eom1}--\eqref{eom3} for the singular field configuration is a complicated problem which can have more than one possible solution. However the process of solving the boundary value problem for the field $\phi$ is greatly simplified if the condition $n=g^2N\ll 1$ is imposed \cite{Son:1995wz}.  Then one needs to find only the initial euclidean part of the solution.  In this case the resulting \emph{saddle point value} of the Euclidean action is
\begin{equation}\label{iz}
\textrm{Im}S[\phi]=S_E[\phi]=\frac{1}{2}e^{-\theta}\int d \textbf k\,a_\textbf k^*a_ \textbf ka^{\omega_\textbf kT}\equiv \frac{1}{2}e^{-\theta}I(T),
\end{equation}
where the last equality defines the function $I(T)$, and $a_\textbf k$ are the Fourier components of the initial field asymptotics \eqref{eom2} rewritten as
\begin{equation}\label{bd3}
\phi(\textbf k)=\frac{a^*_\textbf {k}}{\sqrt{2\omega_k}} e^{-\omega_\textbf k\tau}
\end{equation}
for $\tau=-it\to +\infty$. The saddle point equations \eqref{T} then allow to express the parameter $T$ and $\theta$ in terms of the average energy $\epsilon=E/N$ (here we consider massless particles) and particle number $N$ as 
\begin{equation}\label{T2}
\epsilon=\frac{I'(T)}{I(T)},\quad \theta=-\ln N+\ln I(T).
\end{equation}
Finally, for the scattering cross section we obtain \cite{Son:1995wz,Bezrukov:1999kb}:
\begin{align}\label{cs3}
\sigma(E,N)&=\exp\left(N\ln g^2 N-N+Nf(\epsilon)\right),\\\label{cs4}
f(\epsilon)&=\epsilon\, T(\epsilon)-\ln g^2 I(T),
\end{align}
where by writing $T=T(\epsilon)$ we stress that $T$ should be expressed through $\epsilon$ by solving \eqref{T2}. The energy dependence of the scattering cross section is contained in the function $f(\epsilon)$. 

To summarize, this semiclassical approach allows one to determine the exponent of scattering cross section for multiparticle process $\textrm{few}\to N$, see
\cite{Bezrukov:1999kb}.  To do this one first has to find a set of solutions of the Euclidean equations of motion $\delta S_E/\delta\phi=0$,
singular on the surface $\tau_s(\mathbf{x})\leq0$, $\tau_s(0)=0$ with initial asymptotics \eqref{bd3}.  Then one has to extremize the integral $I(T)$ for some fixed value of $T$ over all values of $a_\textbf k$
(or, equivalently, extremize over the singularity surfaces).  Finally, from the equation \eqref{T2} one obtains the value of $\epsilon$ corresponding to the given $T$ (this is equivalent to extremization over $T$ for given $\epsilon$), and uses \eqref{cs3} and \eqref{cs4} to calculate the cross section. This method applies to any scalar field theory with a dimensionless parameter $g$ such that under the change of variables $\phi=\Phi/g$ the action transforms as in \eqref{con}. Then in the semiclassical limit
\begin{equation}\label{limit}
g^2\to 0,\quad \varepsilon\equiv g^2E=\textrm{fixed},\quad\,n\equiv g^2N=\textrm{fixed}\ll1
\end{equation}
the scattering cross section for the multiparticle process with the total energy $E$ and the particle number $N$ in the final state can be obtained as described above. We note that the condition $n\ll1$ is not essential for the applicability of the saddle point approximation (\ref{cs0})--(\ref{con}).  This condition allowed to simplify the original boundary value problem to solution of only the Euclidean part of the equations (\ref{cs0})--(\ref{con}), leading to the simple prescription described in (\ref{iz})--(\ref{cs4}), see \cite{Son:1995wz}.
The terms of order $\mathcal O (n^2=g^4N^2)$ in the exponent of the scattering cross section arise only from loop corrections.  It means that this approximation is equivalent to considering only the tree-level contribution to the scattering cross section.

Note also, that if we perform extremization over only a subclass of the singularity surfaces (eg.\ only O(4) symmetric ones), then the resulting cross section provides a \emph{lower bound} on the cross section, analogously  to a Rayleigh-Ritz extremization procedure, see \cite{Bezrukov:1999kb} for the detailed proof.

\section{$\lambda\phi^4$ theory}\label{sec:3}
The semiclassical approach to the calculation of the cross section for the process $\mathrm{few}\to N$ for large $N$ was previously applied in \cite{Son:1995wz,Khlebnikov:1992af,Bezrukov:1999kb} to the $\lambda \phi^4$ theory with the action 
\begin{equation}\label{act1}
S=\int d^4x\left(\frac{1}{2}\left(\partial_\mu\phi\right)^2-\frac{\lambda}{4}\phi^4\right).
\end{equation}
In this Lagrangian the coupling constant $\lambda$ is dimensionless and thus the dimensionless saddle point parameter is simply $g^2=\lambda$. Indeed, it is straightforward to check that the action satisfies the condition \eqref{con}. Thus the multiparticle scattering cross section in the limit \eqref{limit} can be evaluated semiclassically by using the formulae \eqref{cs3}, \eqref{cs4} with $g^2=\lambda$. The saddle point value of the Euclidean action \eqref{iz} for the $O(4)$ symmetric case can be found analytically \cite{Son:1995wz,Khlebnikov:1992af}. In the more complicated case of the massive $\lambda\phi^4$ theory the saddle point has been found numerically in \cite{Bezrukov:1999kb}. As expected, in high energy region it was shown to reproduce the results of the massless case.  

\begin{figure}
\begin{center}
\includegraphics[width=10cm]{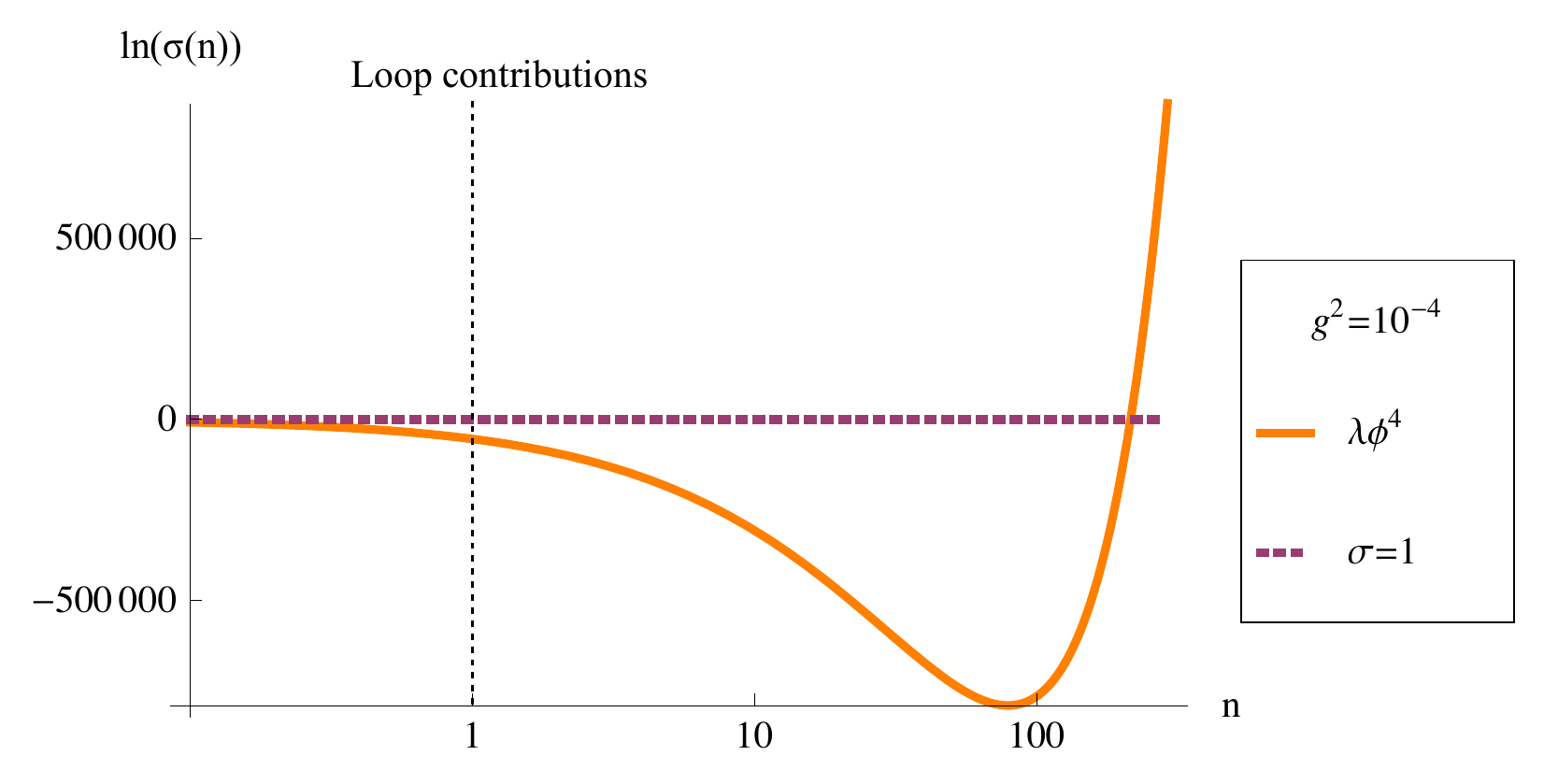}
\caption{\scriptsize{The scattering cross section for multiparticle production in $\lambda\phi^4$ theory depending on the semiclassical particle number $n\equiv \lambda N$, evaluated for $\lambda=g^2=10^{-4}$. For the values $n>1$ the loop contributions have to be taken into account.}}
\label{fig:1}
\end{center}
\end{figure}

In the massless $\lambda\phi^4$ theory the function $f(\epsilon)$, and consequently also the scattering cross section, is independent of energy, i.e.\ it is simply a constant\footnote{This numerical value of the function $f(\epsilon)$ coincides with that given by Son \cite{Son:1995wz}, but might differ from other authors, e.g. \cite{Bezrukov:1999kb}, due to the alternative definition of $f(\epsilon)$ in \eqref{cs4}.} $f(\epsilon)=\ln (1/8\pi^2)$. The scattering cross section as a function of particle number $N$ in the final state for any value of energy $E$ is then
\begin{equation}
\sigma(E,N)=\sigma(N)=\exp\left[N\ln\left(\frac{\lambda N}{8e\pi^2}\right)\right].
\end{equation}
In terms of the semiclassical variable $n\equiv\lambda N$ this becomes
\begin{equation}\label{csl1}
\sigma(n)=\exp\frac{1}{\lambda}\left[n\ln\left(\frac{n}{8e\pi^2}\right)\right].
\end{equation}
Fig.~\ref{fig:1} shows the exponent of the scattering cross section. 

We see that the multiparticle production is exponentially suppressed till the particle number reaches the critical value $n=8e\pi^2\approx215$ above which the exponent in \eqref{csl1} becomes positive. This means that the result obtained in the saddle point approximation cannot be trusted beyond this point. However, we know that the result for the scattering cross section was obtained in the limit $n\ll 1$ and hence the loop contributions become important in the region where the values of the semiclassical variable $n>1$. The positivity of the exponent for the semiclassical particle number values $n>8e\pi^2$ is thus well outside the validity region of our tree level approximation. 

\section{Scalar DBI theory}\label{sec:4}
Let us now consider the following Euclidean DBI type action
\begin{equation}\label{dbi}
S_E=\epsilon_2\int d^4x\,\frac{1}{2L_*^4}\sqrt{1-2 \epsilon _2L_*^4\left(\partial_\mu\phi\right)^2},
\end{equation} 
where all the quantities are dimensionful, i.e.\ $[\phi]=L^{-1}$ and the coupling constant has the dimension of length $[L_*]=L$. The parameter $\epsilon _2$ can take values $\pm 1$. In order to make use of the semiclassical approach described in previous sections one has to introduce a dimensionless parameter, which would play the role of the saddle point expansion parameter $g$. For this we perform the following rescaling of the scalar fields $\phi\to\phi/g$ where the parameter $g$ is arbitrary. The action transforms as
\begin{equation}\label{dbi2}
S_E(\phi/g)=\epsilon_2\frac{1}{g^2}\int d^4 x\,\frac{1}{2l^4}\sqrt{1-2\epsilon_2l^4\left( \partial_{\mu}\phi\right)^2}=\frac{1}{g^2}s(\phi,\,l^4),
\end{equation}
where
\begin{equation}\label{defl}
l^4\equiv \frac{L_*^4}{g^2}.
\end{equation}
We see that the parameter $1/g^2$ factors out in front of the action $s(\phi,\,l^4)$ and the action becomes dependent on the new parameter $l^4$. It is useful to separate the parameters of the theory in two groups: the physical and semiclassical. The physical parameters of the theory are the dimensionful coupling constant $L_*$ and the energy and particle number in the final state $E$ and $N$ respectively. The semiclassical variables were introduced in section \ref{sec:2} as quantities which remain fixed in the semiclassical limit, when $g\to 0$. Besides the semiclassical energy $\varepsilon\equiv g^2E$ and semiclassical particle number $n\equiv g^2N$, in DBI theory there is an additional quantity which has to stay constant in the limit $g\to 0$. We see this from the action \eqref{dbi2} since it explicitly depends on the new parameter $l^4$. It is clear that, in order to evaluate the action in saddle point approximation, also the $l^4$ has to remain fixed. The corresponding limit \eqref{limit}, in which the tree level multiparticle scattering cross section in DBI theory can be evaluated in saddle point approximation, is then
\begin{equation}\label{limit2}
g\to 0,\quad \varepsilon\equiv g^2 E=\textrm{fixed},\quad  l^4\equiv \frac{L_*^4}{g^2}=\textrm{fixed},\quad n\equiv g^2 N=\textrm{fixed}\ll1.
\end{equation} 
The conditions $\varepsilon,\,n,\,l^4=\textrm{fixed}$ define the region of applicability of the saddle point approximation to the scattering problem, whereas the condition $n\ll1$ is needed in order to simplify calculations by neglecting the possible loop contributions.

There are two interesting features of the semiclassical limit of the scalar DBI theory. First we observe that the product $EL_* =\varepsilon l/g^{3/2}$ becomes large in the limit $g\to 0$, corresponding to the interesting case of energies exceeding the cutoff scale, i.e.\ $E>L_*^{-1}$. The second observation is that it is possible to introduce a length scale associated with a given energy such that this length remains constant in the semiclassical limit. Indeed, by setting $r_*(E)=E^\alpha L_*^{1+\alpha}$ and replacing physical variables with the semiclassical ones, we obtain the condition
\begin{equation}
r_*(E)=\varepsilon ^\alpha l^{1+\alpha}g^{-2(1-3\alpha)}=\textrm{fixed} \qquad\Rightarrow\qquad \alpha=\frac{1}{3}.
\end{equation}
This determines the parameter $\alpha$ uniquely and we obtain that $r_*=L_*(EL_*)^{1/3}$. Hence the semiclassical length scale coincides with the classicalization radius introduced in \cite{Dvali:2010jz}.

Let us present the results for the scattering cross section of the process $\textrm{few}\to N$. For simplicity we limit the extremization procedure to the O(4) symmetric singularity surfaces of the classical solution. As we will see, in DBI theory the derivative of the field is singular in distinction from the $\lambda\phi^4$ theory where the field itself was singular. Nevertheless, the previous conditions for the finiteness of the energy \eqref{E2} are still satisfied for the singularity in the first derivative.\footnote{In the truncated DBI theory with only $  L_*^4(\partial_\mu\phi\partial^\mu\phi)^2$ self-interaction term the singularity appears in the second derivative of $\phi$. In order to apply the semiclassical technique to this case the initial state should be chosen as $\left|i\right\rangle=\exp (J\dot\phi(0))\left|0\right\rangle$.} The equation of motion obtained by varying the action \eqref{dbi2}, in terms of the 4-dimensional radial coordinate $\rho$, is
\begin{equation}
\partial_\rho\left[\rho^3\frac{\partial_\rho\phi}{\sqrt{1-2\epsilon_2 l^4(\partial_\rho\phi)^2}}\right]=0,
\end{equation}
and hence
\begin{equation}\label{dr}
\frac{d\phi}{d\rho}=\frac{R_s^3}{\sqrt{2}l ^2}\frac{1}{\sqrt{\rho^6+\epsilon _2R_s^6}}.
\end{equation}
For $\epsilon_2 =-1$ the derivative becomes singular at the singularity radius $\rho=R_s$. In order to obtain the solution, which is singular at the coordinate origin $\tau=|\textbf x|=0$, one has to choose another coordinate system where the Euclidean time coordinate is shifted as $\tau\to\tau+R_s$, so that
\begin{equation}
  \rho^2 = (\tau+R_s)^2+\mathbf{x}^2.
\end{equation}
For $\epsilon_2 =+1$ the derivative is regular everywhere. 
Hence due to the lack of a singular O(4) symmetric Euclidean solution the semiclassical method for calculation of multiparticle scattering cross sections cannot be restricted to this subclass of solutions in this case. Instead, for the $\epsilon_2=+1$ branch of the DBI theories some more generic subclass of singularity surfaces should be considered, which is however beyond the scope of the present work. Henceforth we will therefore investigate the $\epsilon_2=-1$ case.

It is interesting to note that according to a recent paper by Dvali et al. \cite{Dvali:2012zc} the classicalization at all UV energy scales should occur in the $\epsilon_2 =+1$ case. For the $\epsilon_2=-1$ case the classicalization, if present at all, is expected to happen in some finite energy range $E_*<E<\bar\omega$ \cite{Dvali:2011nj,Dvali:2012zc}. The \emph{declassicalization} scale $\bar\omega$ is model dependent and, in general, depends on the scale, at which some new weakly coupled degrees of freedom should be integrated in, and the theory is UV completed in the usual Wilsonian sense.

After setting $\epsilon_2=-1$ the solution of the equation of motion \eqref{dr} for $\phi(\rho)$ is
\begin{equation}
\phi(\rho)=\frac{1}{\sqrt{2}l^2}\int^{\rho}\frac{d\rho'}{\sqrt{\left(\frac{\rho'}{R_s}\right)^6-1}}.
\end{equation}
In the asymptotic region $\rho\to\infty$ the integral can be approximately taken as
\begin{equation}\label{sol1}
\phi(\rho)=\frac{-R_s^3}{\sqrt{2}l^2}\frac{1}{2\rho^2}.
\end{equation}
and the Fourier components have the following asymptotics at $\tau\to\infty$
\begin{equation}\label{four1}
\phi(\tau,\textbf k)=\frac{a_\textbf k^*}{\sqrt{2\omega_\textbf k}}e^{-\omega_\textbf k\tau},\quad \textrm{where }\omega_\textbf k=|\textbf k|\textrm{  and  }a_\textbf k^*=\frac{R_s^3}{2l^2}\sqrt{\frac{\pi}{2\omega_\textbf k}}e^{-\omega_\textbf kR_s}.
\end{equation}

\begin{figure}
\begin{center}
\includegraphics[width=10cm]{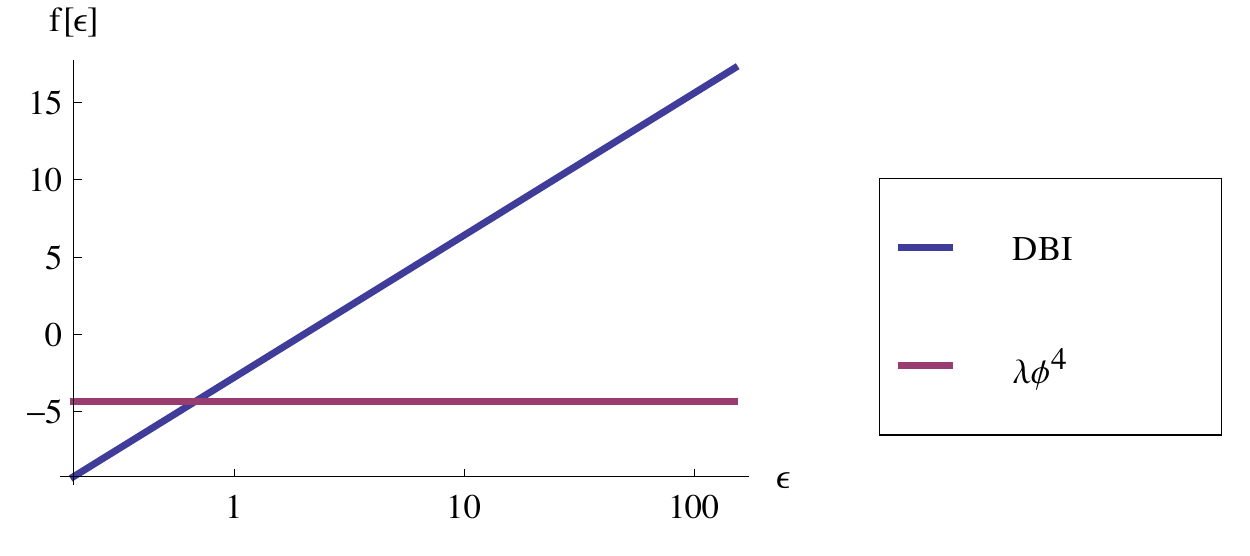}
\caption{\scriptsize{Function $f(\epsilon)$ for $\lambda\phi^4$, and DBI theory evaluated at $g^2=\lambda=10^{-4}$.}}
\label{fig:dbi}
\end{center}
\end{figure}

The saddle point value of the Euclidean action $I(T)$ in \eqref{iz} is then
\begin{equation}
I(T)=\frac{R_s^6}{L_*^4}\frac{\pi^2}{2}\frac{1}{(2R_s-T)^2}.
\end{equation}
After extremizing the function $I(T)$ over all $R_s$ we obtain for the function $f(\epsilon)$ the following expression
\begin{equation}\label{fe}
f(\epsilon)=4+\ln\frac{2^3}{\pi^23^6}+4\ln(\epsilon),
\end{equation}
where we have used $g^2=L_*^4/l^4$. In distinction from the $\lambda\phi^4$ case this function grows with the energy density $\epsilon\equiv E/N$ as shown in Fig.~\ref{fig:dbi}. 

After substituting this expression of $f(\epsilon)$ in \eqref{cs3} we find the tree level scattering cross section
\begin{equation}\label{cs5}
\sigma(E,N)=\exp\left[3N\ln\frac{N_{\textrm{crit}}}{N}\right]=\left(\frac{N_\textrm{crit}}{N}\right)^{3N},
\end{equation}
where we have defined the \emph{critical particle number} in the final state $N_\textrm{crit}$ as
\begin{equation}\label{ncrit}
N_\textrm{crit}^3\equiv c^3(L_* E)^4,\quad c^3=\frac{(2e)^3}{\pi^23^6}.
\end{equation}
We see that for fixed total energy $E$ the scattering process $\textrm{few}\to N$ is only suppressed for $N>N_\textrm{crit}$. The notion of critical particle number allows one to define a \emph{critical length scale} such that for given energy $E$ it satisfies
\begin{equation}\label{rcrit}
r^{-1}_\textrm{crit}\equiv\frac{E}{N_\textrm{crit}},\quad\Rightarrow\quad r_\textrm{crit}=cL_*( L_* E)^{1/3}.
\end{equation}
In other words, $r_\textrm{crit}^{-1}$ corresponds to the maximal allowed energy per particle and coincides with the classicalization radius $r_*$ defined in \cite{Dvali:2010jz,Dvali:2010ns}. Hence we have shown that the classicalization radius $r_*$ emerges as the critical length scale at which the behavior of the scattering cross section drastically changes. 

We will discuss the behavior of the transition rate in dependence of the particle number $N$ in the final state for fixed energy $E$ in two separate energy regions. 

\subsection{Strong coupling region: $E>L_*^{-1}$}
\begin{figure}
\begin{center}
\includegraphics[width=12cm]{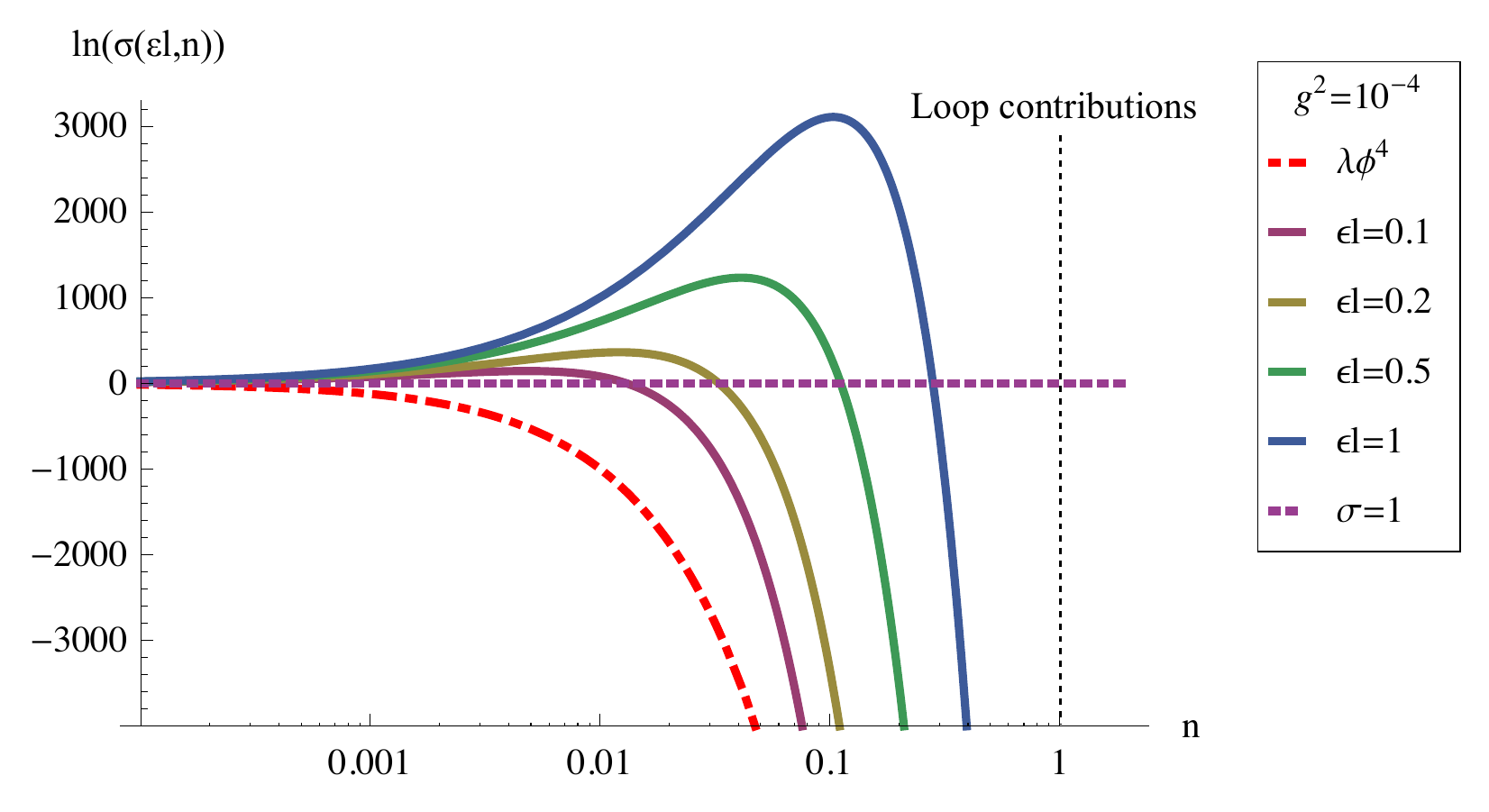}
\caption{\scriptsize{The exponent of the scattering cross section as a function of the semiclassical particle number in the final state $n$ for DBI theory with $\epsilon_2=-1$ evaluated at different values of the parameters $\varepsilon l =EL_* g^{3/2}$. The numerical value of the semiclassical parameter $g^2=10^{-4}$ and hence the parameter region $\varepsilon l >g^{3/2}=10^{-3}$ corresponds to the high energy region $EL_*>1$. For the values $n>1$ the loop contributions have to be taken into account.}}
\label{fig:cs2}
\end{center}
\end{figure}

It is useful to rewrite the expression for the scattering cross section \eqref{cs5} in terms of the semiclassical variables defined above
\begin{equation}\label{cs6}
\sigma(\varepsilon,n)=\exp\frac{1}{g^2}\left[3n\ln\left(\frac{c(\varepsilon l)^{4/3}}{n}\right)\right].
\end{equation}
We see that the functional dependence of the cross section is very different from the $\lambda\phi^4$ theory in equation \eqref{csl1}. In $\lambda\phi^4$ theory the scattering is exponentially suppressed for small values of the semiclassical particle number $n$. Meanwhile in DBI theory the exponent of the scattering cross section becomes \emph{positive} for small values of $n<c(\varepsilon l)^{4/3}$ and thus the expression \eqref{cs6} cannot be trusted for these values of $n$. The comparison of the dependence of the scattering cross section on the semiclassical particle number $n$ in DBI theory and in $\lambda\phi^4$ theory is shown in Fig.~\ref{fig:cs2}.

We recall here that a similar break-down of the saddle point approximation is observed in $\lambda\phi^4$ theory for large values of $n$. However that is an artifact of the tree level approximation $n\ll 1$ since we have neglected all terms of order $O(n^2)$. The behavior of the scattering cross section at larger values of $n$ is changed by the loop contributions \cite{Son:1995wz}. The same logic also applies to the DBI theory, but as is shown in Fig.~\ref{fig:cs2} the higher order corrections become relevant only at values $n>1$. Hence the break-down of the semiclassical approach cannot be cured by adding higher order corrections to the exponent of the scattering cross section \eqref{cs6}. 

\begin{figure}
\begin{center}
\includegraphics[width=12cm]{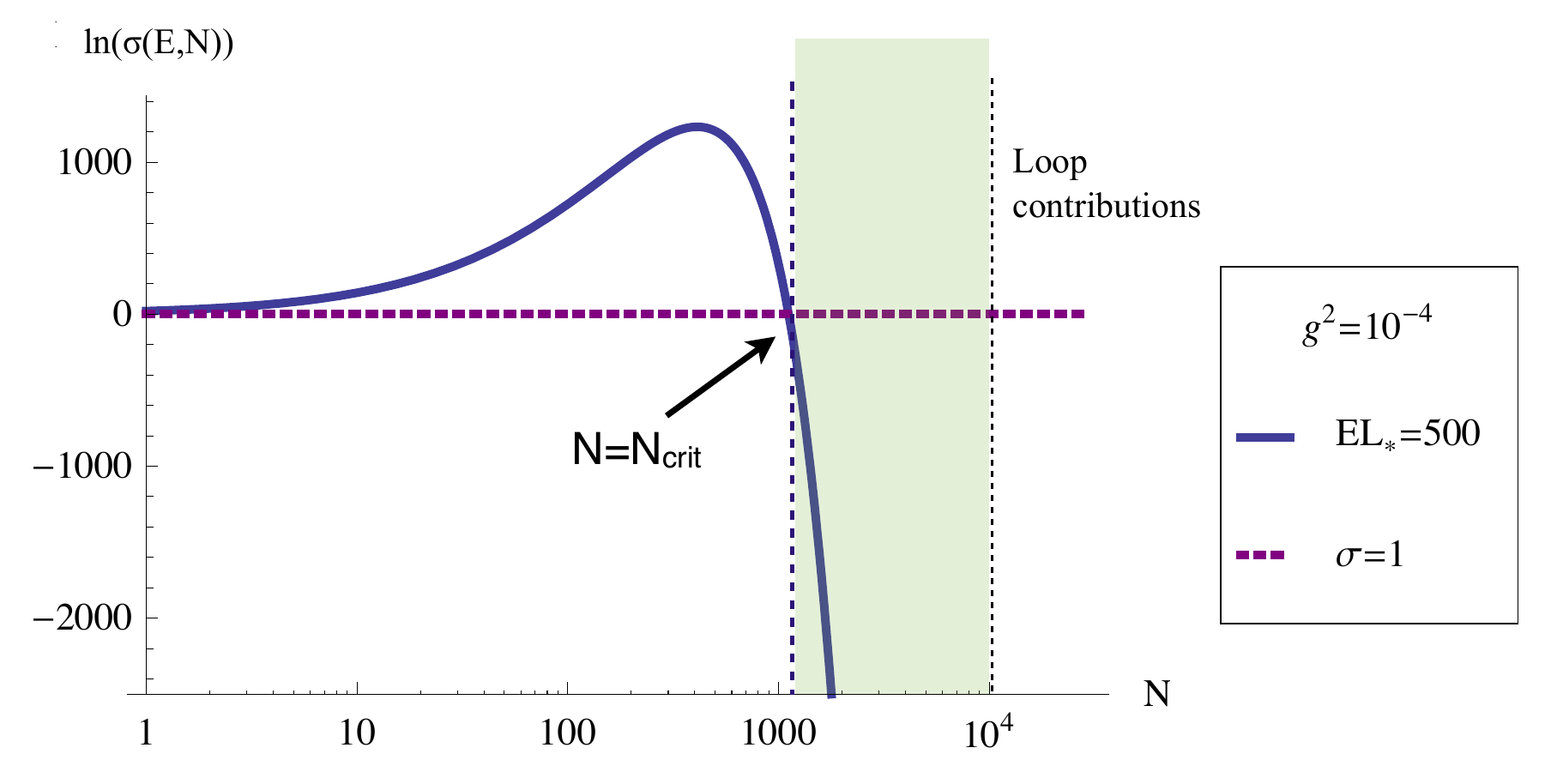}
\caption{\scriptsize{The exponent of the scattering cross section as a function of the physical particle number in the final state $N$ for DBI theory with $\epsilon_2=-1$ evaluated at energy $EL_*=500\gg1$ } The numerical value of the semiclassical parameter $g^2=10^{-4}$. The saddle point method breaks down for $N<N_\textrm{crit}=c(EL_*)^{4/3}$ while the loop contributions become important at $g^2N\sim 1$. The region in which the semiclassical method gives a reliable result is painted in green.}
\label{fig:4}
\end{center}
\end{figure}

In terms of the physical particle number it means that the semiclassical method does not allow to make conclusive statements about the scattering cross sections for the procesess where few initial particles scatter into $N<N_\textrm{crit}=c(EL_*)^{4/3}$ particles with the total energy $E>L_*^{-1}$. Remarkably, the saddle point method gives a reliable result for the transition rates to final states with particle number larger than the critical. In this region the scattering processes are exponentially suppressed. The scattering cross section as a function of the physical particle number in the final state is shown in Fig.~\ref{fig:4}. We note that with perturbative methods this energy region is completely unaccessible. It is therefore interesting to find that there exists a region for large particle number in the final state $N>N_\textrm{crit}$ where the non-renormalizable theory behaves semiclassically. The energy per particle in the final state of $N_\textrm{crit}$ particles equals $r_\textrm{crit}^{-1}=\left[cL_*( L_* E)^{1/3}\right]^{-1}\ll L_*^{-1}$ and thus the final state is composed of very soft particles as suggested by classicalization arguments in \cite{Dvali:2010jz,Dvali:2010ns}. For even larger particle number the energy per particle further decreases. 

\subsection{Perturbative region: $E\lesssim L_*^{-1}$}
The region of the physical particle numbers where the exponent of the scattering cross section is positive has no physical meaning as soon as the critical particle number $N_\textrm{crit}=c(EL_*)^{4/3}$ becomes less than one. In this case the scattering process is exponentially suppressed for all physically reasonable values of the particle number in the final state $N>1$. This happens for low energies $(EL_*)<c^{-3/4}=2.57$. Strictly speaking this requirement translates into a condition on the semiclassical parameters $EL_*=\varepsilon l/g^{3/2}<c^{-3/4}$ which is not satisfied in the semiclassical limit when $g\to 0$ (however, for some numerically small values of the parameters $g$ and $\varepsilon l$ the condition can still be fulfilled). Nevertheless, the obtained result is physically reasonable since the exponent of the scattering process is negative. Hence formally the semiclassical method can also be applied for the energy values which are below the non-renormalizability cutoff. However the obtained results should be compared with results from perturbative calculations. The perturbative check for the exponentiation of the scattering amplitude for $2\to N$ transitions in $\lambda\phi^4$ theory was done in \cite{Libanov:1995gh,Libanov:1994ug}. The same procedure could be applied also to DBI theory. 

\section{Conclusions}\label{sec:5}
We have applied the semiclassical approach to the calculation of the scattering cross sections for the multiparticle production from a few-particle initial state in a classicalizing theory. A reliable result is obtained in two parameter regions for the energy $E$ and particle number $N$ in the final state. First, the exponential suppression is observed below energy cutoff $E\lesssim L_*^{-1}$ for any number of particles $N>1$. This corresponds to the parameter region also accessible with perturbative methods. The second range of parameters leading to trustable results lies above energy cutoff  $E>L_*^{-1}$ but is restricted to large particle numbers $N>N_{\textrm{crit}}$ only. This result is obtained in the region where the theory is strongly coupled and perturbation theory cannot be used. No information about hard high energy scattering processes $\textrm{few}\,\to\,N<N_{\textrm{crit}}$ is obtained from the semiclassical approach.

Let us discuss, how robust is the failure of the applied semiclassical procedure at $E>L_*^{-1}$, $N<N_{\textrm{crit}}$.  First, we limited the analysis here to the O(4)-symmetric singularity surfaces, while at least in the case of the $\lambda\phi^4$ theory it is known, that the true extremum of the boundary value problem is reached on a generic surface \cite{Bezrukov:1999kb}.  However, this in general should lead to even larger cross sections, and thus should not resolve the break-down of the saddle point approximation. Another option can be that the O(4)-symmetric family of the semiclassical solutions passed through a bifurcation point at the typical energy $E\sim L_*^{-1}$, making it an irrelevant subclass of the classical solutions at high energies.  Another promising reason may be related to the fact that the limit of the vanishing source $j\to0$, leading to the singular solutions, no longer commutes properly with the semiclassical limit and is not imitating a few particle initial state (the conjecture is checked by explicit comparison with the perturbation theory only in the normal renormalizable theories, see \cite{Libanov:1995gh,Libanov:1996vq}).  In this regard an approach, alternative to using the initial expression (\ref{cs0}), may prove valuable.  However, previous attempts to get a real time classical solution, corresponding to the high energy spherical collisions, lead to development of singularity at high energies \cite{Heisenberg:1952zz,daniel,Rizos:2011wj,Brouzakis:2011zs}. Hence it is not clear, if non-singular relevant semiclassical solutions exist.  Thus, further study of the real time solutions is needed to get a useful insight into the classicalization phenomena.

We therefore do not have conclusive statements about the presence or absence of the classicalization phenomena since this demands a better understanding of hard scattering processes with a small particle number in both initial and final state. The critical behavior of the trans cutoff multiparticle production was observed at the number of particles which corresponds to $N_{\textrm{crit}}$ very soft particles with the energy per particle given as $r_*^{-1}=\left[L_*(EL_*)^{1/3}\right]^{-1}$. This coincides with the inverse of the classicalization radius introduced in \cite{Dvali:2010jz,Dvali:2010ns}. With this we have shown the emergence of this critical length scale in the semiclassical approach which is conceptually completely different from the classical perturbative estimates of \cite{Dvali:2010ns}.

\section{Acknowledgments}
The authors would like to thank Gia Dvali, Andrei Khmelnitsky, Dima Levkov, Maxim Libanov, and Sergey Troitsky for helpful and enlightening discussions and comments on the draft. LA is supported by the DFG Cluster of Excellence EXC 153 ``Origin and Structure of the Universe''.

\end{document}